\documentclass[11pt,a4paper]{article}
\usepackage[utf8]{inputenc}
\usepackage{amsmath}
\usepackage{amsfonts}
\usepackage{amssymb}
\usepackage{graphicx}
\usepackage{bm}
\usepackage{epstopdf}
\usepackage{hyperref}       
\usepackage{url}            
\usepackage[top=2.52cm, bottom=2.52cm, left=2.52cm, right=2.52cm]{geometry}
\usepackage{xcolor} 
\begin{document}
\title{\textbf{Nanoporous alumina microtubes for metamaterial and plasmonic applications}}



\author{Dheeraj Pratap\footnote{dheerajpratap.sulr@gmail.com}, S. Anantha Ramakrishna \\
Department of Physics, Indian Institute of Technology Kanpur, Kanpur - 208016, India}

\date{}

\maketitle

\begin{abstract}
Double anodization of aluminium microwires in an acid bath yields cylindrical nanoporous alumina with nonbranching radially emanating pores.
The obtained nanoporous alumina is a cylindrically anisotropic as well as a radially inhomogeneous optical medium. Detailed structural characterization reveals that the nanopore diameter varies linearly with the radius of the aluminium microwire along the radial direction.  Microcracks form on the alumina shell during the anodization when sufficient thickness is formed due to volume expansion and stress accumulation. The formation of the microcracks can be monitored by the anodization current which shows sudden jumps when the cracks are formed. After removing the remaining aluminium at the core of the anodised wire the anisotropic and inhomogeneous alumina microtube is obtained. Such nanoporous alumina microtubes form unique optical waveguides and are useful for microscale heat transfer applications. 

\vspace{3mm}
\textbf{Keywords:} Metamaterial, Nanoporous alumia, Microtube, Anisotropic, Plasmonic, Inhomogeneous. 
\end{abstract}


\section{Introduction}
Anodization of aluminium sheets is well known for several decades for fabricating nanoporous alumina with high aspect ratio pores~\cite{masuda1995,Li1998,Lee2007b}. A double anodization process, where the nanoporous alumina formed in the first anodization process is etched off to yield a textured aluminium surface followed by a second anodization process to again form the alumina, results in highly ordered non-branching and straight nanopores organized on a hexagonal lattice~\cite{masuda1995}. A variety of acids can be used for the anodization and result in different nanopore sizes and interpore separations~\cite{sousa2014}. The nanopore diameter as well as the structural composition of the alumina can be also controlled by the anodization voltage~\cite{virk2008, Santos2014}. Anodization at low temperatures of the acid bath results in more ordered arrays of nanopores~\cite{Sellarajan2016}. Subsequent chemical etching can enlarge the pore sizes to any desired value up to almost the interpore separation distances~\cite{Zhao2007}. The possibility of structural modification of nanopores in the nanoporous alumina makes it a very flexible system for creating templates to deposit other materials~\cite{Yao2008, Sharma2007,Shingubara2003}. This technique has been also used to make highly anisotropic nanowire metamaterials~\cite{Yao2008} where the nanowires are organized parallel to each other.

Nanoporous alumina in the cylindrical geometry has been rarely reported~\cite{Niwa2002,Mizushima2004,Chae2005} and have been used for chemical catalysis~\cite{Niwa2002,Mizushima2004}. Alumina coated aluminium needles for use as low friction hypodermic needles has also been reported~\cite{Chae2005}. In previous reports, only alumina membrane tubes with large diameters of several hundred micrometers have been developed. We adopted the anodization method, which is well known in the literature for creating planar layers of nanoporous alumina,~\cite{masuda1995,Li1998,Lee2007b} to create a cylindrically symmetric system. The anodization of cylindrical aluminium wires with diameters of sub-hundred micrometers, and their detailed structural characterization and analysis are not well known. Nanoporous alumina in the cylindrical geometry has very different properties from the usual nanoporous alumina in their structural and anodization characteristics. Recently, we reported these as the manifestation of an anisotropic metamaterial fiber and discussed the optical properties of these fibers~\cite{pratap2015,pollock2016,pratap2018}. There a theoretical study of the mode structures and preliminary optical measurements have been reported~\cite{pratap2015}. But these reports lack the details of the synthesis of the nanoporous alumina microtubes, and their structural characterization and analysis.
\begin{figure*}[h!]
\begin{center}
\includegraphics[width=12cm]{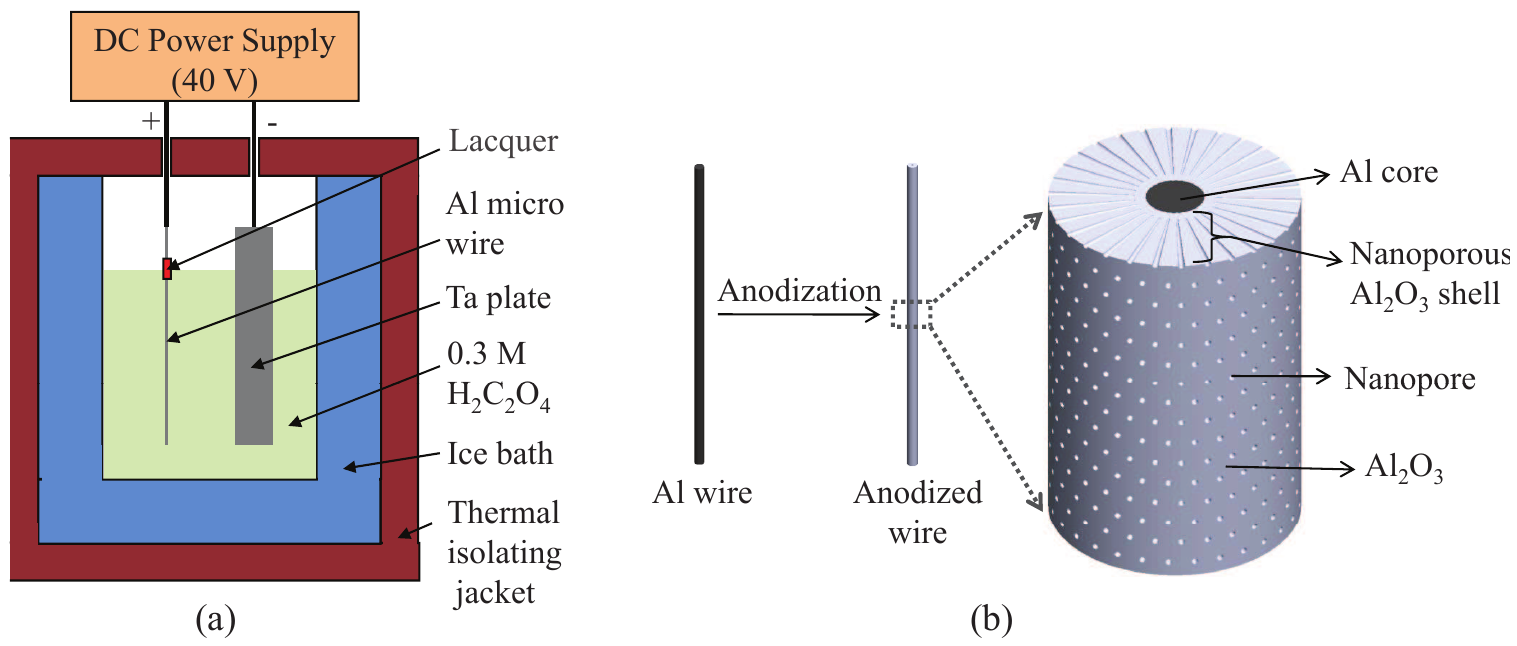}
\end{center}
\caption 
{ \label{fig:anodization_schematic}
Schematic pictures of (a) the setup to anodize an Al microwire and (b) the details of the structure of the anodized wire.} 
\end{figure*}

In this article, we discuss in detail the experimental synthesis and characterization of these nanoporous alumina microtubes. The nanopores have a cross-section that changes along with the radial distance, thereby rendering the structurally anisotropic microtube, an inhomogenous system as well. The radial inhomogeneity in the fabricated microtubes and the effect on the effective material parameters are characterized. In section - 2, we first present the experimental details of the synthesis of the nanoporous alumina microtubes. In section - 3, structural characterization and analysis are given. In the final section - 4, we conclude with a discussion of applications as cylindrical metamaterials. 
\section{Synthesis of nanoporous alumina microtubes}
A schematic of the setup used for the anodization of an aluminium microwire is shown in Fig.~\ref{fig:anodization_schematic}(a). Aluminium microwires (99.999 \% purity, Alfa Aesar) of the circular cross-section of diameter 100 $\mu$m were first annealed in vacuum at 500 $\mathrm{^o}$C for three hours to release any stresses as well as to enlarge the grain sizes.  After degreasing with acetone, the wires were electropolished in a  mixture of ethanol (C$_2$H$_5$OH) and perchloric acid (HClO$_4$)  in the volume proportion of 5:1 at a current density of 0.48 A/cm$^2$ for three minutes. At the solution-air interface, lacquer was applied on the microwire of aluminium to protect it from the fast reaction due to the meniscus formed at the interface (Fig.~\ref{fig:anodization_schematic}) and a double anodization process was carried out on the wire.  A tantalum (Ta) plate was used as the counter electrode for the anodization while an ice bath was used to keep the temperature constant at 0 $\mathrm{^o}$C. The first anodization of the electropolished microwire was carried out using 0.3 M oxalic acid (H$_2$C$_2$O$_4$) in de-ionized (DI) water solution at 40 V positive potential applied for 30 minutes. As even small turbulence in the solution bent the thin wires, the solution was not stirred.  After first anodization, the nanoporous alumina (Al$_2$O$_3$) formed was etched off in a mixture of chromic oxide (CrO$_3$)  and phosphoric acid (H$_3$PO$_4$) in DI water at 90 $\mathrm{^o}$C  temperature for 45 minutes. The second anodization of alumina etched microwires of aluminium was carried out at the same parameters as for the first anodization for 2 hours to 12 hours to result in different thicknesses. After second anodization the aluminium core was removed in a mixture of cupric chloride (CuCl$_2$), hydrochloric acid (HCl) and DI water at room temperature. The aluminium core could be removed from the anodized wire for lengths of few centimeters.  Longer etching times that are required to etch out longer lengths resulted in partial dissolution and damage to the alumina nanoporous structure.

In Fig.~\ref{fig:anodization_schematic}(b) the schematic picture of the anodized wire shows the details of the structures. There is a nanoporous alumina cylindrical shell and the remaining (un-anodized) aluminium is a core at the center. The nanopores in the nanoporous alumina shell are radially oriented.  At the outer surface of the anodized wire, the nanopores are ideally organized in a hexagonal pattern. Note that the process gives us the full flexibility to obtain different pore sizes, interpore distances by parameters such as the nature of the acid. The interpore separation at the initial aluminium surface depends on the voltage used and the acid~\cite{virk2008,sousa2014,Santos2014,Sellarajan2016}. The size of the nanopores can be reduced (or increased) while anodization process by the physical and/or chemical parameters. Keeping chemical parameters fixed, and decreasing the voltage~\cite{sacco2018investigation} or temperature~\cite{aghili2019fabrication} the pore size can be decreased. Keeping physical parameters the same and by lowering the concentration of the electrolyte the pore dimension can be reduced~\cite{voon2016effect,pashchanka2016evidence}. By starting with a thicker aluminium microwire and etching off the alumina formed, one can obtain a microwire of a given thickness with a textured surface having a high density of nucleation centers for the formation of the nanopores upon further anodization. This is due to the radially conversing pores. Since the porous alumina is brittle and the system is a very thin microtube, there is a need for great care to handle such delicate systems, which can be done by micromanipulators. 
\section{Characterization and analysis of nanoporous alumina microtubes}
\begin{figure*}[h]
\begin{center}
\begin{tabular}{c}
\includegraphics[width=12cm]{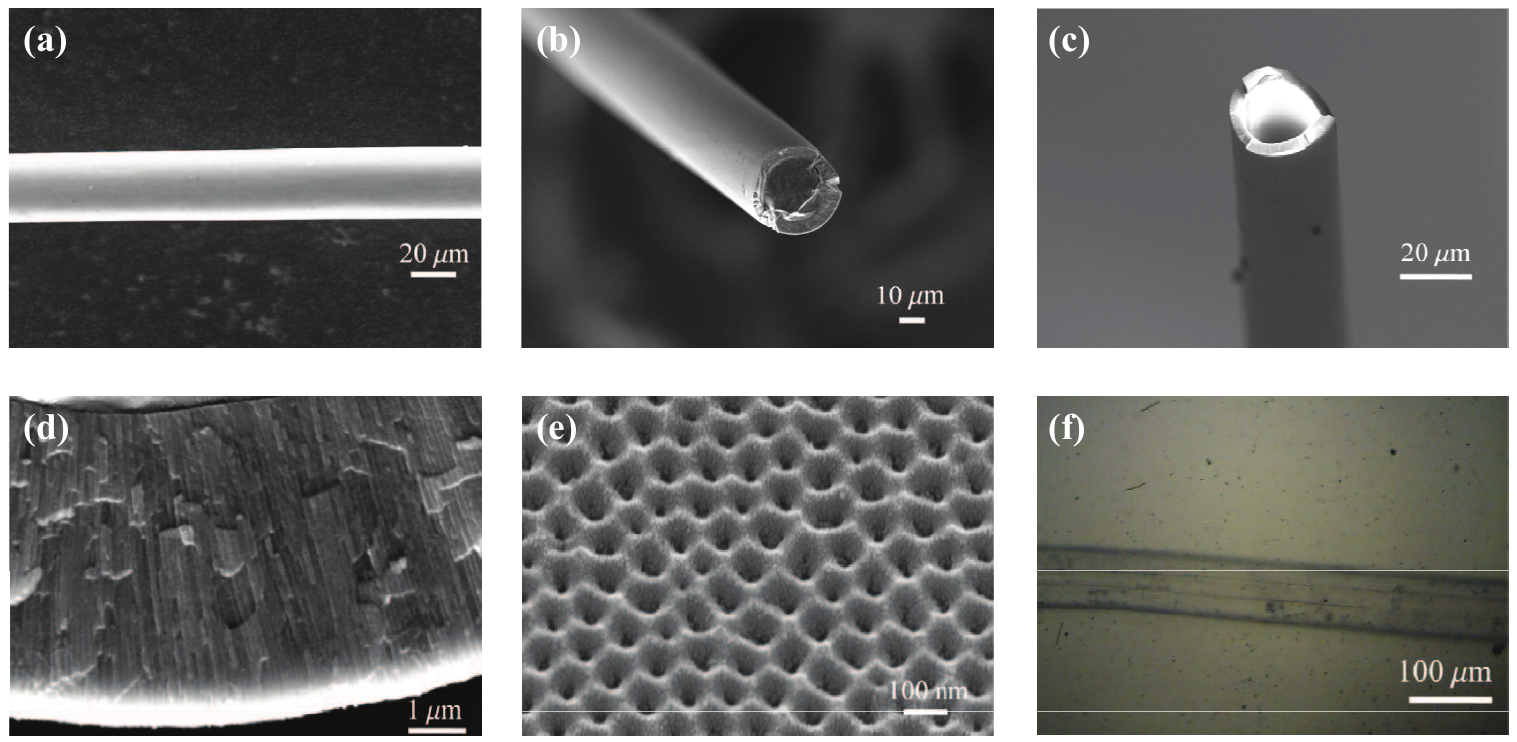} 
\end{tabular}
\end{center}
\caption 
{ \label{fig:sem_image}
FESEM images: (a) an electropolished aluminium microwire, (b) anodized aluminium wire with aluminium core, (c) cylindrical nanoporous alumina microtube with removed aluminium core, (d) cross-section of the cylindrical nanoporous alumina, and (e) the outer surface of the anodized aluminium wire. (f) An optical microscope image of the nanoporous alumina shell demonstrates the transparency of the microtube.} 
\end{figure*}
The structural characterization of the cylindrical nanoporous alumina microtube was carried out by a field emission scanning electron microscope (FESEM - Zeiss Sigma). A FESEM image of the smooth surface of an electropolished aluminium microwire is shown in Fig.~\ref{fig:sem_image}(a). Since the anodization proceeds perpendicular to the surface~\cite{Jessensky1998}, nanopores are oriented along the radial direction, and consequently form a cylindrical nanoporous alumina shell. Figure~\ref{fig:sem_image}(b) shows the cross-section of the anodized aluminium microwire with an aluminium core and a nanoporous alumina shell. The alumina shell thickness increases approximately linearly with the anodization time at a constant voltage. Due to the geometric singularity at the \textit{z}-axis of the cylinder the aluminium microwires can not be completely anodized up to center and a metallic core will always be left behind. A nanoporous alumina microtube with the aluminium core removed by dissolving out the aluminium using CuCl$_2$ is shown in  Fig.~\ref{fig:sem_image}(c). The radial orientation of the nanopores towards the center of the microwire is seen from the cross-section in Fig.~\ref{fig:sem_image}(d). In Fig.~\ref{fig:sem_image}(e), the outer porous surface of the anodized microwire is shown to have a nanopore diameter of 30 nm and an inter pore distance of 100 nm. The porosity and pore density of the nanoporous alumina using the FESEM images are obtained 8\% and $2 \times 10^{10}$ /cm$^2$ are quite close to the reported values in literature~\cite{Jessensky1998,lee2006fast}. The surface area and porosity of the nanoporous alumina using the Brunauer-Emmet-Teller (BET) analysis were reported as 6 m$^2$/g and 0.016 cm$^3$/g respectively~\cite{ibrayev2014plasmon,rui2014anodic}. These values of surface area and porosity of the nanoporous alumina using the BET analysis are very small as compared to the surface area (231 m$^2$/g) and porosity (0.23 cm$^3$/g) of the mesoporous alumina as reported by A . K. Patra \textit{et al.}~\cite{patra2012self}. An optical microscope image of the hollow core nanoporous alumina microtube (Fig.~\ref{fig:sem_image}(f)) demonstrates the optical transparency of the microtube, wherein the inner air core can be delineated. In general, for many optical uses the nanoporous alumina microtube should be transparent. The X-ray diffraction (XRD) data of the nanoporous alumina shows an amorphous structure~\cite{masuda2015fabrication,vera2020understanding}. The crystal structure of the nanoporous alumina changes from amorphous to crystalline form by thermal treatment~\cite{vera2020understanding,choudhari2012fabrication}.

\begin{figure*}[h]
\begin{center}
\begin{tabular}{c}
\includegraphics[width=12cm]{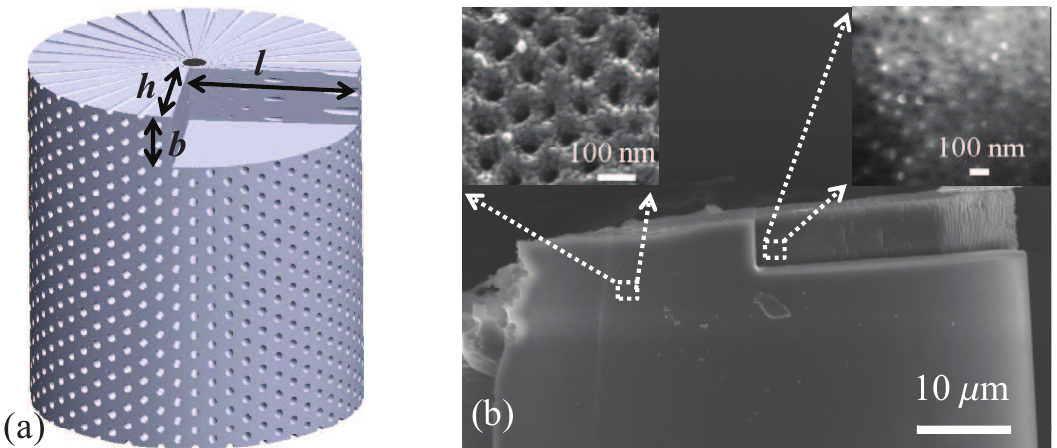} 
\end{tabular}
\end{center}
\caption 
{ \label{fig:depth_effect}
(a) The schematic for FIB cutting the cylindrical nanoporous alumina with length $l$ (22 $\mu$m), width $b$ (4 $\mu$m) and depth $h$ (12 $\mu$m) and (b) the FESEM image of the pores of the FIB cut nanoporous alumina microtube at the outer surface and depth 12 $\mu$m.} 
\end{figure*}
To clearly understand the variation of the nanopores along the radial direction, a volume of 22 $\times$ 4 $\times$ 12 $\mu$m$^3$ has been milled out from the microtube of diameter 45 $\mu$m  using a focused ion beam (FIB – FEI Nova 600) as shown schematically in Fig.~\ref{fig:depth_effect}(a). In Fig.~\ref{fig:depth_effect}(b), the FESEM images show the nanopore cross-section at the outer surface and 12 $\mu$m depth inside the microtube along the radial direction. Because of the depth and liquid-metal ion source (LMIS) etching, the image of the exposed inner surface is not as sharp due to defocus as the image at the outer surface.  But it gives clear evidence that the pore diameter decreases with the radial distance towards the axis of the anodized wire and further the areal density of the nanopores is increased. Essentially each nanopore has a conical geometry with a small cone angle. We note that despite the breaking of symmetry along with the axial and azimuthal directions, the cross-section of the nanopores remains circular that minimizing the free energy. The radially oriented nanopores make the system structurally anisotropic while varying diameters of the nanopores make the system inhomogeneous. The dielectric properties of the metamaterial will now clearly have a variation with the radial distance. 

Figure~\ref{fig:depth_effect} shows that the size of the nanopore decreases radially towards the center of the microtube. If we equate the solid angle of nanopore (diameter $=Q$) at the outer surface to the solid angle of nanopore (diameter $=q$) at the inner surface of the nanoporous alumina shell we get the size of the nanopore (and hence the fill fraction of nanopore) is a function of the radial position, $q=(Q/R)r$ where $R$ is the outer diameter of the shell and $r \leq R$. The nanoporous alumina microtubes allow fine control of the nanostructural morphology which can be used to define the effective medium parameters such as dielectric permittivity. The nanoporous alumina microtube system has anisotropy of $\varepsilon_r \neq \varepsilon_\phi = \varepsilon_z$~\cite{pratap2015} rather than $\varepsilon_r = \varepsilon_\phi \neq \varepsilon_z$~\cite{russell2003,tuniz2013}. The latter anisotropy is shown by photonic crystal fibers and drawn metamaterial fiber which are structured along the axis. An increase of the inhomogeneity by controlling the nanopore diameter with radial distance by change of anodization parameters during the anodization process and pore widening can enable fine control over the inhomogeneity and anisotropy, and hence, over the optical properties. Using this system one can	create radially increasing or decreasing optical inhomogeneity (effective refractive index) depending on the material filled in the nanopores. For a fixed dimension of the nanoporous alumina microtube, the size of the nanopore plays an important role in the degree of anisotropy or inhomogeneity of the cylindrical nanoporous alumina which can be controlled while anodization process~\cite{sacco2018investigation,aghili2019fabrication,voon2016effect,pashchanka2016evidence}.
\subsection{Effect of radius of curvature: Anodization current and crack formation}
\begin{figure*}[h]
\begin{center}
\begin{tabular}{c}
\includegraphics[width=12cm]{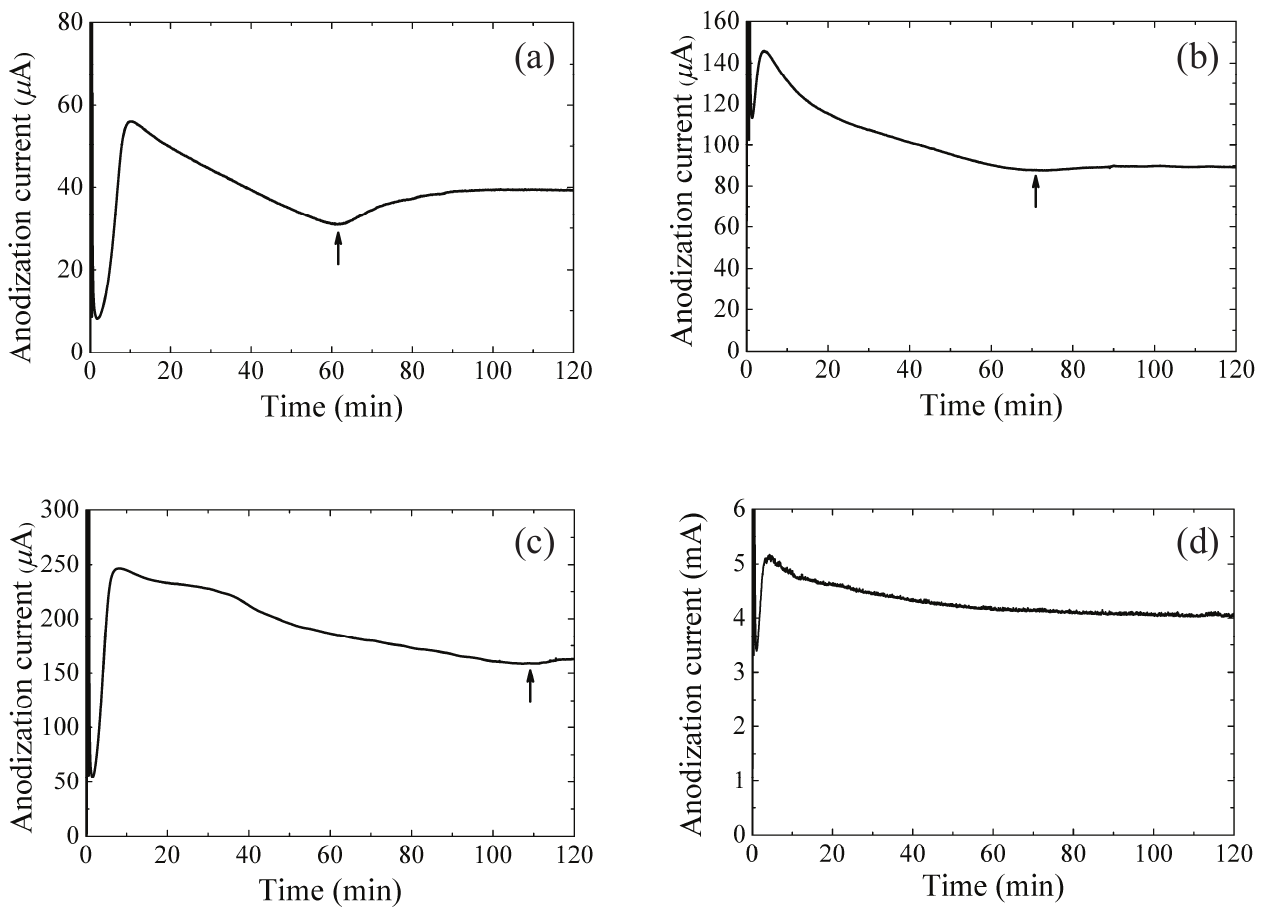} 
\end{tabular}
\end{center}
\caption 
{ \label{fig:anodization_current}
The measured anodization current of aluminium microwires of diameters (a) 50 $\mu$m, (b) 100 $\mu$m and (c) 220 $\mu$m. The small jump in the anodization current of anodization of aluminium microwires of (a), (b) and (c) occurred around 60 minutes, 70 minutes and 110 minutes respectively. (d) The anodization current of an aluminium sheet for reference. There is no jump in the anodization current of aluminium sheet anodization.} 
\end{figure*}
We noted that during the anodization of aluminium microwires, a crack could develop in the nanoporous alumina shell after some time.  Comparatively, there is usually no crack in the planar nanoporous alumina systems. We recorded the anodization current of aluminium microwires of diameters 50 $\mu$m, 100 $\mu$m, 220 $\mu$m, and an aluminium sheet. The recorded anodization currents are shown in Fig.~\ref{fig:anodization_current}. The anodization current in the case of the aluminium microwires first increases quickly to a peak value and then decreases over time.  While decreasing, a sudden jump occurs in the
\begin{figure*}[h!]
\begin{center}
\includegraphics[width=12cm]{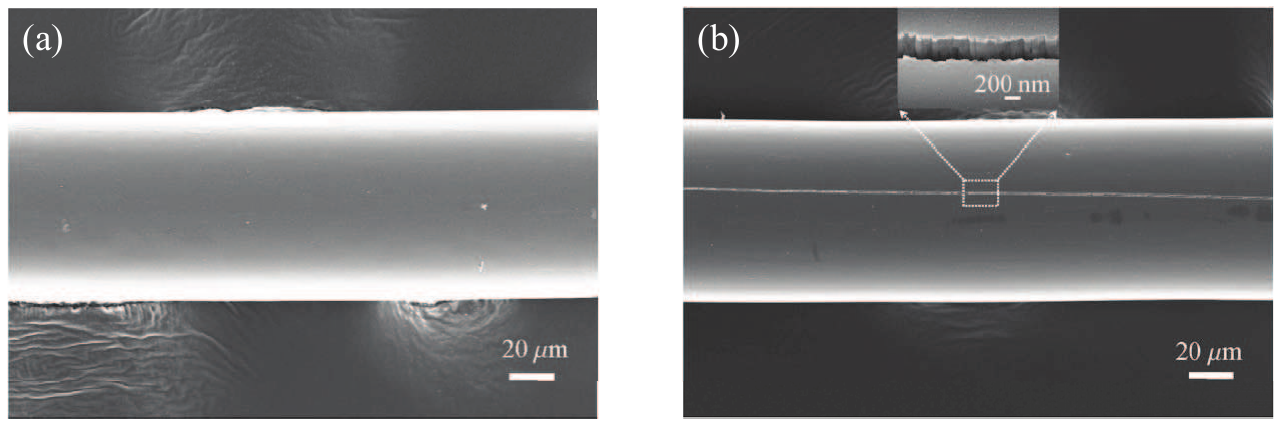} 
\end{center}
\caption 
{ \label{fig:crack}
FESEM images of anodized aluminium wire (a) 30 min anodization time which show no crack and (b) 70 min anodization time just after the jump in the anodization current occurred.} 
\end{figure*}
 anodization current beyond the peak value after some time and the anodization current then decreases again. The times at which the jumps in the anodization current occurred for microwires of 50 $\mu$m, 100 $\mu$m and 220 $\mu$m diameters were around 60 minutes, 70 minutes and 110 minutes respectively. This is shown in sub-figs.~\ref{fig:anodization_current}(a), \ref{fig:anodization_current}(b) and \ref{fig:anodization_current}(c) respectively (marked by the arrows). In comparison, the anodization current of the aluminium sheet after the peak value always decreases over time~\cite{Li1998,Lee2007b,Li2012} and here there is no jump in the anodization current as shown in Fig.~\ref{fig:anodization_current}(d). This jump in the microwire anodization current occurs because of the formation of the microcracks. When the crack occurs the effective resistivity between solution and aluminium microwire decreases and the anodization current slightly increases. For larger diameters of the microwires, the time at which the current changes rapidly shifts towards longer times. For the planar aluminium sheet, the radius of curvature is effectively infinite and the anodization current continues to steadily decrease with no sudden changes. 

\begin{figure*}[h!]
\begin{center}
\includegraphics[width=12cm]{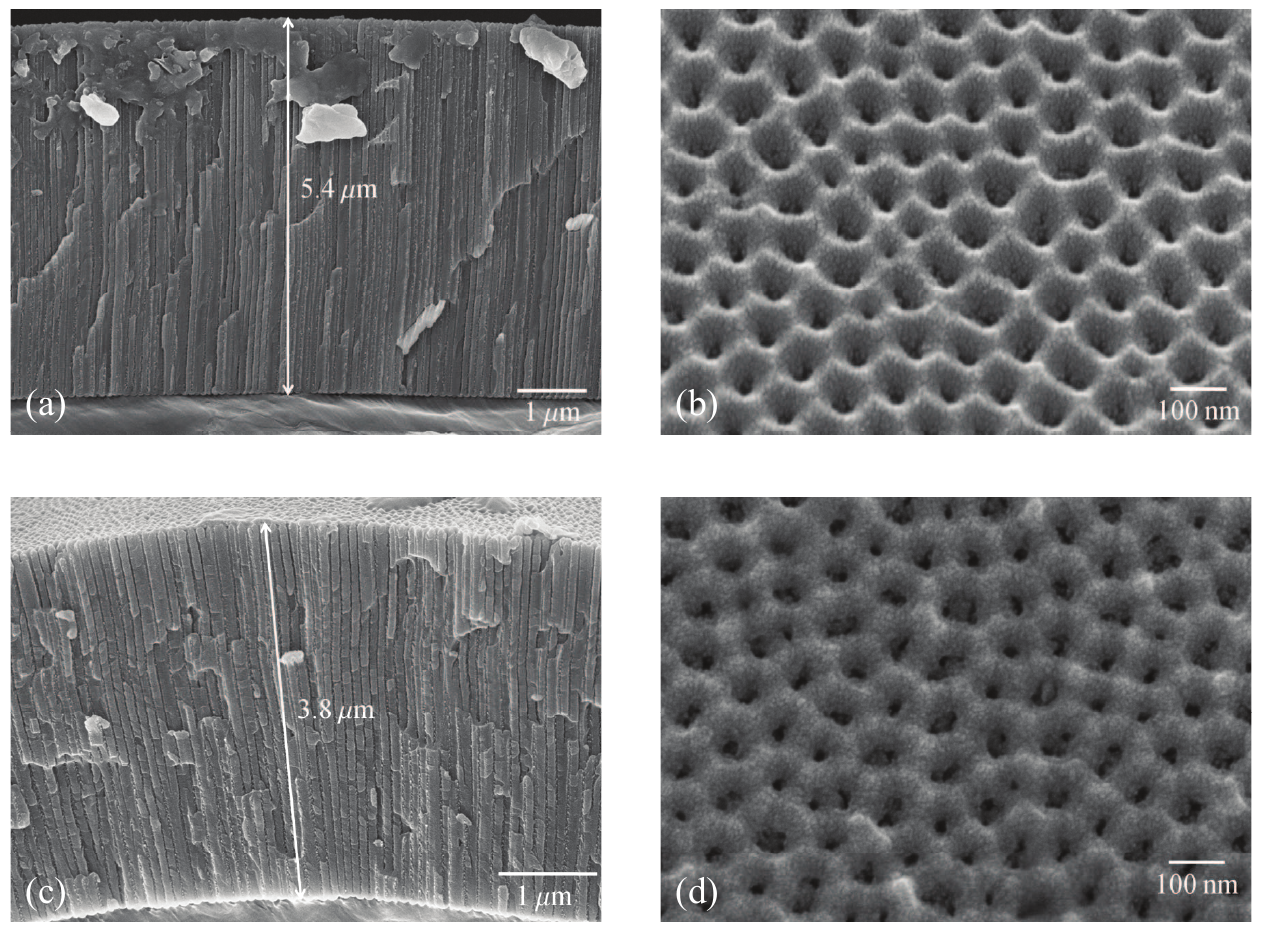} 
\end{center}
\caption 
{ \label{fig:pore_organization}
FESEM images of anodized aluminium microwire anodized for two hours.  (a) and (b) are the cross-section and outer surface of an anodized wire of 220 $\mu$m diameter respectively. The anodized thickness is 5.4 $\mu$m, the average pore diameter at the outer surface is 30 nm and inter pore distance is 100 nm. (c) and (d) are the cross-section and outer surface of an anodized wire of  25 $\mu$m diameter. Here the anodized thickness is 3.8 $\mu$m, average pore diameter and inter pore distance are 30 nm and 100 nm respectively.} 
\end{figure*}
In Fig.~\ref{fig:crack}, two anodized wires of the same diameter have been shown. Figure~\ref{fig:crack}(a) shows the anodized microwire after 30 minutes of anodization and there is no crack in the microtube. Figure~\ref{fig:crack}(b) shows the same microwire after 70 minutes of anodization. The sample had been taken out of the acid bath immediately after the jump in the anodization current occurred. We found that a small crack in the nanoporous alumina shell had appeared. If the aluminium microwire is anodized for a longer time, then the crack size also becomes larger and more cracks may also form. We know that the anodization proceeds perpendicular to the planar aluminium sheet surface and there is a volume expansion associated, as the volume of nanoporous alumina is much larger compared to the original volume of the aluminium~\cite{Jessensky1998}. In the case of the planar aluminium sheet anodization, the reactive area of aluminium always remains constant throughout the anodization while the volume of nanoporous alumina increases perpendicular to the sheet surface. So the cracks do not occur in the planar nanoporous alumina because its reactive area is always constant. In the case of anodization of the aluminium microwire, the anodization starts perpendicular to the curved surface along the radial direction towards the center in the cylindrical symmetry. So the effective reactive area continuously decreases with radius and the relative volume expansion of the nanoporous alumina increases. Thus, the alumina is under tensile stress and also plastically deforms to some extent. Eventually, with continued build up stress, a crack suddenly appears as the yield point is crossed. For the smaller diameters of the microwire, the effective reactive area decreases faster compared to the larger diameter microwire, so the crack formation occurs much earlier for microwires with smaller diameters. 
\subsection{Effect of radius of curvature: Organization of nanopores and shell thickness}
Now we investigate the pore organization for different starting thickness of the aluminium microwires.  Two aluminium microwires of diameters 220 $\mu$m and 25 $\mu$m respectively were anodized for 2 hours. The FESEM images of these anodized aluminium wires are shown in Fig.~\ref{fig:pore_organization}. The nanoporous alumina thickness on the  220 $\mu$m wire is 5.4 $\mu$m and on the  25 $\mu$m thick wire is 3.8 $\mu$m. As the radius of the wire decreases, the pore diameter has to decrease fast, thereby reducing the surface area on the semi-permeable alumina layer for the reaction. Thus, the reaction rates reduce and the thickness of the nanoporous alumina formed decreases for the same anodization time. Figure~\ref{fig:pore_organization} also shows the top surfaces of these two anodized wires where we see that the average inter pore distance is 100 nm and the average pore diameter is 30 nm on the outer surface for both the anodized wires. At the beginning of the anodization, the pore nucleation centers form at an optimum distance and hence the inter pore distance on the outer surface is the same.  The angular spacing of the nanopores thereby is smaller for a wire with an initial larger thickness. The interpore separation is principally determined by the specific acid used for the anodization, the anodization voltage and bath temperature.~\cite{virk2008,sousa2014} For example, the use of oxalic acid at 0 $^{\circ}$C and 40 V for anodization typically yields our structure with interpore separation of about 100 nm and pore diameter of 30 nm. Pore widening is easily carried out by the chemical action of dilute phosphoric acid and it is independent of the diameter of the aluminium wire. Hence the starting thickness of the aluminium wire determines the angular spacing of the nanopores for a given acid, which will be small for thicker wires and large for thinner wires. If a thin nanoporous alumina microwire is desired with smaller angular spacing, one can start by anodizing a thicker wire, etch off the alumina formed leaving behind a thinner wire but a pre-textured surface that might lead to a smaller angular pore spacing (as for the thicker wire). Thus, any given pore volume fraction can be attained through this process and pore widening by etching.    
\subsection{Embedding the alumina microtube with plasmonic nanoparticles}
\begin{figure*}[h!]
\begin{center}
\begin{tabular}{c}
\includegraphics[width=12cm]{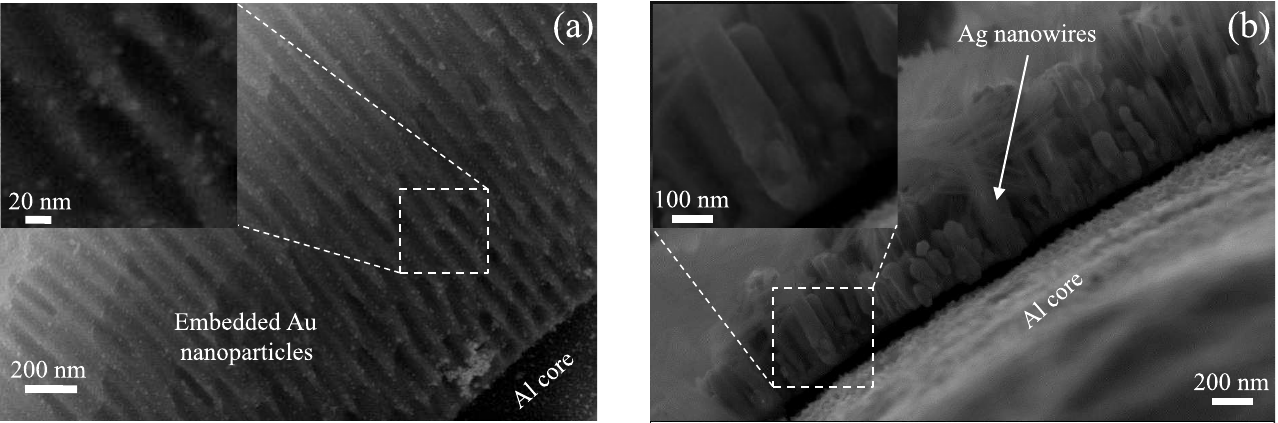} 
\end{tabular}
\end{center}
\caption 
{ \label{fig:embedded}
FESEM images of cross-section of the nanoporous alumina microtube embedded with (a) Au nanoparticles and (b) Ag nanowires.} 
\end{figure*}
In this section, we demonstrate how plasmonic nanoparticles can be embedded within the nanopores of the alumina microtube easily. This is important from several contexts such as functionalizing the microtubes, using plasmonic nanoparticles as sensors etc. After the anodization process,  nanopore widening of nanoporous alumina was carried out in diluted phosphoric acid for 20 minutes at room temperature and rinsed in DI water and dried on a hot plate at 50 $^{\circ}$C to remove the moisture in the nanopores. 

We first demonstrate embedding the nanopores with gold (Au) nanoparticles. Synthesis of the Au nanoparticles was carried out by a standard process proposed by Jana \textit{et al.}~\cite{Jana2001}. To make Au nanoparticles, 0.1 ml of HAuCl$_4$ (0.25 mM) and 0.150 ml of Na$_3$C$_6$H$_5$O$_7$ (0.25 mM) were made to total 20 ml aqueous solution in a conical flask. Further, while stirring a 0.05 ml of ice-cold, freshly prepared  NaBH$_4$ (0.1 M) solution was added to the solution. After adding NaBH$_4$ the solution turned pink immediately indicating the formation of Au nanoparticles. The size of the nanoparticles was estimated by the absorption spectra, which indicated an average size of about 5 nm and was later confirmed by electron microscopy. Now the pore widened nanoporous alumina was immersed in the suspension of the Au nanoparticles for several hours while periodically shaking it. The microtube was taken out and dried. Figure~\ref{fig:embedded}(a) shows the FESEM images of the cross-section of the alumina microtube embedded with Au nanoparticles. The zoomed image of the Au nanoparticles inside the nanopores is shown in the inset of the Fig.~\ref{fig:embedded}(a). It is seen that the gold nanoparticles pervade the entire thickness of the nanoporous alumina shell and can even be found literally at the bottom of the nanopores. Thus, a reasonably uniform distribution of nanoparticles within the nanopores is achieved. The elemental confirmation of the Au nanoparticles inside the nanopores of the nanoporous alumina was carried out by the Energy-dispersive X-ray spectroscopy (EDAX). Since the average size of the Au nanoparticles was 5 nm, therefore, in the EDAX analysis its weightage percent was very small about 5.83\% compared to the host aluminium and oxygen.

We next demonstrate the possibility of silver (Ag) nanowires within the nanopores of the alumina microtube by electrodeposition. The aluminium core at the center of the alumina shell is used as the cathode and a platinum wire used as the anode. To remove the barrier layer of impermeable alumina inside the cylindrical nanoporous alumina on the aluminium, the voltage after anodization was reduced slowly to zero in steps of 5 V from 40 V to 5 V, and thereafter in steps of 1 V/minute. Subsequently, the templates were immersed in 5\% phosphoric acid for 30 minutes to remove the barrier layer completely. Electrodeposition was performed in  an ice-cold mixture of AgNO$_3$(3.5 gm.), CH$_3$COONH$_4$(2.8 gm.), NH$_3$H$_2$O (25\% extra pure) (4.61 ml.) in 70 ml of solution~\cite{Wang2002} with 3 mA current for 40 minutes. Figure~\ref{fig:embedded}(b) shows the FESEM images of alumina microtube embedded with the electrodeposited Ag nanowires within the nanopores of a porous alumina shell. Inset of the Fig.~\ref{fig:embedded}(b) shows the magnified image of the Ag nanowires. The average diameter of the Ag nanowire was 88 nm. The elemental confirmation of Ag was carried out by EDAX which showed a high weightage per cent about 47.58 \% compared to the host aluminium and oxygen elements. Here, the shell thickness of the nanoporous alumina microtube is, however, small compared to the diameter of the aluminium wire. As the shell thickness increases, the nanopore size decreases towards the center and it becomes very difficult to remove the barrier layer. 
Some other methods, such as a  cathodic polarization~\cite{Zhao2007}, might prove successful to remove the barrier layer from the much thicker nanoporous alumina shell. Further work to optimize the procedures for barrier layer removal and electrodeposition is required.

Metallized nanoporous alumina microtubes with metal nanoparticles or nanowires can be used for the metamaterials and plasmonic applications. The concept of metamaterial is based on effective medium parameters for structured materials with subwavelength structures~\cite{ramakrishna2008}. The properties of the nanoporous alumina microtubes can be very flexibly controlled by filling the nanopores of the nanoporous alumina with desired and suitable materials like metal, nanoparticles etc. We have demonstrated the deposition of gold nanoparticles and silver nanowires inside the nanopores of the porous alumina microtube. The structure of this system is at the subwavelength scale from infra-red to visible wavelengths, where this system can be used for various applications. Theoretical studies show that cylindrical nanoporous alumina microtubes exhibit hyperbolic dispersions after filling silver nanowires in the nanopores. So it can be considered as a cylindrical metamaterial~\cite{pratap2015,pollock2016}. Fine control and optimization of the anodization parameters should enable a better quality of the nanoporous alumina microtubes in future and would be very useful for plasmonic, photonic and metamaterial applications~\cite{ramakrishna2008,capolino2009} such as low order whispering gallery modes, optical hyperlens and metamaterial fiber. The anodised microwires and hollow alumina microtubes also have unique applications in heat transfer at micron size length scales~\cite{arya2016,Anantha2016}. The application to optical waveguiding structures poses a challenge as these microcracks are responsible for a large amount of scattering. Similarly, applications as microscale heat transfer devices also require the elimination of the microcracks which threaten the integrity of the microtube. Thus, the optimization of the electrolytic conditions whereby the differential stress is minimized and the development of the microcracks is avoided is an essential step to the further development of applications of this system.
\section{Conclusions}
This article reports the fabrication of cylindrically anisotropic and inhomogeneous nanoporous alumina microtubes. Details of the fabrication of such nanoporous alumina microtubes by anodization techniques have been presented. The anodization of the cylindrical geometry brings in new features due to the curvature. The generation of microcracks in the cylindrical system arises due to differential expansion of the alumina at different radius, which is an effect principally of the curvature of the geometry. The reported effects for cylindrical nanoporous alumina are not known for the planar nanoporous alumina templates. The nanoporous alumina microtubes are anisotropic because of their structured nature. The structural characterization of the microtubes shows that the nanoporous alumina shells are inhomogeneous. The metallization of the microtubes has been shown by embedding the metal nanoparticles and nanowires in the nanopores for the metamaterials and plasmonic applications.  
\section*{Acknowledgments}
SAR acknowledges the Department of Science and Technology, Ministry of Science and Technology (DST) (Project no. DST/SJF/PSA-01/2011-2012) and DP thanks  CSIR - India for fellowship. 



\bibliography{references}   
\bibliographystyle{unsrt}

\end{document}